\RequirePackage{etex}

\documentclass{iau}

\usepackage{amsmath}
\usepackage{graphicx}
\usepackage{multirow}

\begin{document}

\lefttitle{I. Jim\'enez-Serra}
\righttitle{Chemistry in the Galactic Center}

\jnlPage{1}{7}
\jnlDoiYr{2021}
\doival{10.1017/xxxxx}

\aopheadtitle{Proceedings IAU Symposium}
\editors{C. Sterken,  J. Hearnshaw \&  D. Valls-Gabaud, eds.}

\title{Chemistry in the Galactic Center}

\author{Izaskun Jim\'enez-Serra1}
\affiliation{Centro de Astrobiolog\'{\i}a (CAB), CSIC-INTA, Ctra. de Ajalvir km 4, E--28850, Torrej\'on de Ardoz, Spain}

\begin{abstract}
Gas and dust in the Galactic Center are subjected to energetic processing by intense UV radiation fields, widespread shocks, enhanced rates of cosmic-rays and X-rays, and strong magnetic fields. The Giant Molecular Clouds in the Galactic Center present a rich chemistry in a wide variety of chemical compounds, some of which are prebiotic. We have conducted unbiased, ultrasensitive and broadband spectral surveys toward the G+0.693-0.027 molecular cloud located in the Galactic Center, which have yielded the discovery of new complex organic molecules proposed as precursors of the $"$building blocks$"$ of life. I will review our current understanding of the chemistry in Galactic Center molecular clouds, and summarize the recent detections toward G+0.693-0.027 of key precursors of prebiotic chemistry. All this suggests that the ISM is an important source of prebiotic material that could have contributed to the process of the origin of life on Earth and elsewhere in the Universe.
\end{abstract}


\maketitle

\section{Introduction}

The Galactic Center (GC) covers the inner $\sim$600 pc of the Milky Way and it is the closest galactic nucleus to us. The majority of the molecular gas is concentrated in a chain of Giant Molecular Clouds (GMCs) forming the so-called Central Molecular Zone (or CMZ), which appears distributed between galactic longitudes 1.7$^\circ$ and --0.7$^\circ$, and galactic latitudes --0.2$^\circ$ and +0.2$^\circ$. The CMZ accounts for 5\% of the total molecular gas reservoir of the Milky Way ($\sim$3-5$\times$10$^7$$\,$M$_\odot$), but it hosts more than 80\% of the dense gas found in the Galaxy. As a result, the physical conditions of the molecular gas in the GMCs in the Galactic Center are very different from those found in GMCs in the Galactic disk. 

The dense gas in the CMZ has typical H$_2$ volume densities of a few 10$^4$ cm$^{-3}$, which are almost two orders of magnitude higher than those found in the Galactic disk \citep{guesten1983,walmsley1986,mills2018}. In addition, the molecular gas in the CMZ also presents high gas temperatures \citep[$\sim$70-150$\,$K;][]{guesten1985,mills2013,ginsburg2016,krieger2017,zeng2018}, which constrast the low dust temperatures measured in the CMZ \citep[$\leq$20-25 K; see e.g.][]{rodriguez2000,etxaluze2013}. The molecular gas in the Galactic Center is also affected by intense magnetic fields \citep{crutcher1996,pillai2015}, enhanced cosmic-rays ionization rates \citep[$\zeta$$\sim$10$^{-15}$-10$^{-14}$ s$^{-1}$;][]{goto2013,goto2014,lepetit2016,oka2019}, strong ultraviolet (UV) background radiation fields \citep[G$_0$$\sim$10$^3$-10$^4$ Hab;][]{lis2001,goicoechea2004,clark2013}, X-ray flares coming from the central black hole in Sgr A$^*$ \citep{terrier2010,terrier2018}, and low-velocity shocks (v$_s$$\sim$10-50 km s$^{-1}$) produced by cloud shearing and compression by the central bar potential \citep{guesten1980,martin-pintado1997}, which induce high levels of turbulence \citep{shetty2012,kauffmann2017,henshaw2019}. The CMZ is therefore an extreme environment where all these physical processes drive a very peculiar chemistry characterized by the energetic processing of the molecular gas and dust. In the following, we describe the chemical studies that have been carried out in the past decade, to investigate the origin of the rich and complex organic chemistry observed in the Galactic Center.


\section{Chemical studies carried out toward the Galactic Center}
\label{sec:chemstudies}

Millimeter spectral surveys carried out with the 22-m Mopra radio telescope in Australia, imaged the molecular line emission toward the CMZ in the 3mm and 7mm bands \citep{jones2012,jones2013}. Given the limited angular resolution ($\sim$40-65$"$) and sensitivity of these surveys, only abundant molecular species such as HCN, HNC, HCO$^+$, N$_2$H$^+$, HNCO, SiO, or HC$_3$N, together with some of their isotopologues, were detected and found to  be widespread across hundreds of parsecs in the CMZ. These images revealed a similar spatial distribution for the different molecules, and from these data it became clear that the molecular emission in the CMZ experiences sub-thermal excitation produced by the low H$_2$ volume densities of the gas of a few 10$^4$ cm$^{-3}$ \citep{jones2012,jones2013}. These surveys, however, provided little information about the physical processes responsible for the chemistry observed in the Galactic Center. 

To that aim, higher sensitivity observations focusing on individual GMCs were carried out toward the Sgr A$^*$ and Sgr B2 complexes by \citet{harada2015} and \citet{armijos2019,armijos2020}. By using IRAM 30m and APEX data, \citet{harada2015} analyzed a selected set of molecular line transitions taken toward a position in the southwest lobe of the circumnuclear disk (CND) around Sgr A$^*$. By comparing these data with astrochemical models \citep[the UCLCHEM code;][]{holdship2017}, \citet{harada2015} found that the models that best reproduce the IRAM 30m and APEX observations favour a chemistry dominated by shocks with v$_s$$>$40 km s$^{-1}$ and by an enhanced cosmic-ray ionization rate ($\zeta$$>$10$^{-15}$ s$^{-1}$). The analysis of the high-J transitions of CO measured toward the CND with the {\it Herschel Observatory} show that additional effects in the chemistry of the CND may be caused by UV-photon chemistry \citep{goicoechea2013,goicoechea2018}.    

Energetic processing of the molecular gas is also found toward the Sgr B2 GMC. By using ion molecules such as HCO$^+$, HOC$^+$ and CO$^+$ together with the radical HCO,  \citet{armijos2020} established that UV-photochemistry is dominant toward HII regions (as seen by the clear enhancement of CO$^+$), while for the rest of positions analyzed within the molecular envelope of Sgr B2, the observed chemistry could be explained by the combination of shocks and an enhanced cosmic-ray ionization rate, as proposed for Sgr A$^*$. X-ray chemistry does not seem to play a significant role in the chemistry of the dense gas toward Sgr A$^*$ or Sgr B2, because the estimated X-ray ionization rates are too low to have an effect on the chemistry. Therefore, the chemistry observed toward the Galactic Center is driven mainly by shocks and cosmic-rays and, occasionally, by UV photons toward certain locations in the CMZ \citep[see][]{martin2008}. 

From all this, one would think that this energetic processing of the gas and dust would hamper the emergence of chemical complexity in such an extreme environment. However, as shown in the following sections, the Galactic Center not only is an efficient factory of large molecules such as Complex Organic Molecules (or COMs)\footnote{COMs are defined as carbon-based compounds with at least 6 atoms in their molecular structure \citep{herbst2009}.}, but it may provide the key to our cosmic origins. 


\section{The Central Molecular Zone as a COM factory}

The Galactic Center is one of the most prolific regions for the discovery of COMs in the ISM. This is due to the fact that the Galactic Center hosts large amounts of molecular gas that translates into large column densities of material in the line-of-sight. This increases the possibility of detection of low abundance molecules such as COMs. For this reason, the Galactic Center, and in particular the Sgr B2(N) and (M) massive star-forming regions (which harbour several massive hot cores), have been the target of numerous spectral surveys at centimeter, millimeter, sub-millimeter and far-IR wavelengths \citep[see e.g.][]{cummins1986,cernicharo1988,nummelin1998,nummelin2000,friedel2004,belloche2013,moeller2021}. Many COMs have been detected for the first time toward these massive hot cores, with some of the COM species being of prebiotic interest such as amino acetonitrile \citep[NH$_2$CH$_2$CN;][]{belloche2008}, iso-propyl cyanide \citep[i-C$_3$H$_7$CN;][]{belloche2014}, or urea \citep[NH$_2$CONH$_2$;][]{belloche2019}. For the last two molecules, the high-angular resolution images, together with the high sensitivity and imaging fidelity provided by the Atacama Large Millimeter/Submillimeter Array (ALMA), were crucial for the robust identification of these COM species. 

In contrast to what is typical found in GMCs in the Galactic disk where COM emission is found localized in regions of star formation \citep[as e.g. massive hot cores, low-mass hot corinos, or molecular outflows;][]{jorgensen2016,codella2017,oya2017,belloche2019}, COMs in Galactic Center GMCs are widespread across the CMZ \citep{requena2006, jones2011,li2017,li2020}. Interestingly, most of these GMCs are quiescent in the sense that they do not show any signs of star formation and thus, the detection of large COMs in Galactic Center GMCs is attributed to the presence of widespread low-velocity shocks that erode/sputter dust grains injecting vast amounts of COM material into the gas phase \citep{requena2006}. 

By carrying out a COMs survey toward a large sample of Galactic Center GMCs, \citet{requena2006} found that the abundance ratio of COMs with respect to methanol is rather uniform across the CMZ. Recurrent widespread shocks with dynamical time-scales of $\sim$10$^5$ yrs and shock velocities $\geq$6 km s$^{-1}$ are needed in Galactic Center GMCs to explain such uniform COM abundance ratio \citep{requena2006}. Widespread SiO has been found in the CMZ, which supports the idea of large-scale and widely-distributed low-velocity shocks in the center of the Galaxy \citep{martin-pintado1997}. The COM abundance ratios in Galactic Center GMCs were also found to be very similar to those measured in Galactic disk hot cores and hot corinos, which led \citet{requena2006} to suggest that dust grains present a universal ice chemical composition across the Galaxy. 

Galactic Center GMCs not only present large abundances of COMs but they also show a wide variety of them \citep{requena2006,requena2008,widicus2017,zeng2018}. They have indeed been the target of multiple discoveries of COMs such as acetic acid \citep[CH$_3$COOH;][]{mehringer1997,remijan2002}, glycolaldehyde \citep[CH$_2$OHCHO;][]{hollis2000}, ethylene Glycol \citep[(CH$_2$OH)$_2$][]{hollis2002}, 
propenal (CH$_2$CHCHO) and propanal \citep[CH$_3$CH$_2$CHO;][]{hollis2004}, cyclopropenone \citep[c-H$_2$C$_3$O;][]{hollis2006a}, acetamide \citep[CH$_3$CONH$_2$;][]{hollis2006b},   cyanoformaldehyde \citep[CNCHO;][]{remijan2008} and propylene oxide (CH$_3$CHCH$_2$O), the first quiral molecule detected in the ISM \citep[][]{mcguire2016}.
Most of these COMs were detected at centimeter wavelengths in absorption against the bright radiocontinuum emission of Sgr B2 (N) using the Green Bank Telescope (GBT). The reason for seeing these molecules in absorption is because they are present in the Sgr B2 molecular envelope, which is cooler and has lower densities than the bright massive hot molecular cores in Sgr B2 (N) and (M). Therefore, Galactic Center GMCs are one of the richest repositories of COMs in the Galaxy \citep{requena2006,requena2008,widicus2017,zeng2018}.

\section{The Galactic Center molecular cloud G+0.693-0.027}

Among Galactic Center GMCs, the molecular cloud G+0.693-0.027 (hereafter, G+0.693) stands out because it is very rich in COMs of all chemical families: O-bearing COMs \citep{requena2006,requena2008,widicus2017}, N-bearing COMs \citep{zeng2018,rivilla2022c}, S-bearing COMs \citep{rodriguez-almeida2021a}, P-bearing species \citep{rivilla2018,rivilla2022b} and C-chains \citep{fatima2023}. G+0.693 appears as the brightest position in HNCO emission in the single-dish maps of the CMZ obtained by \citet{jones2011} and \citet{jones2012}. It has been proposed that the origin of the chemical richness of the G+0.693 molecular cloud is a cloud-cloud collision, which induces low-velocity shocks that sputter the icy mantles of dust grains \citep{zeng2020}. The emission of Class I CH$_3$OH masers, which are collisionally pumped in shocked gas \citep{kurtz2004,voronkov2006,cyganowski2009}, is indeed extended across the Sgr B2 molecular envelope and peaks close to the location of G+0.693 \citep[see black cross in Figure \ref{fig:survey} and][]{Liechti1996}.    

  \begin{figure}
  \begin{center}
  \includegraphics[scale=0.5,angle=0]{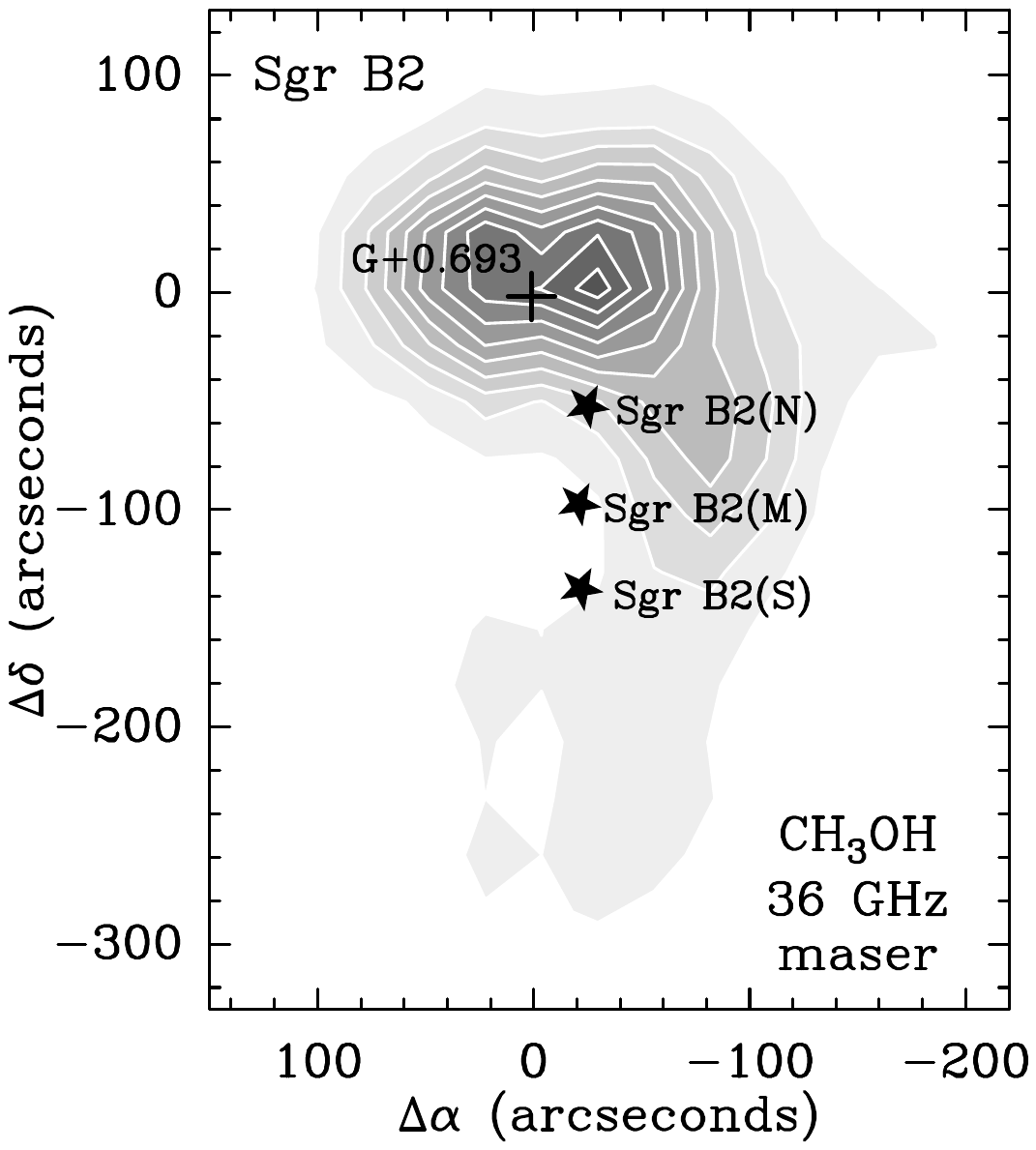}
  \caption{Integrated intensity map of the Class I CH$_3$OH maser at 36 GHz obtained with the Yebes 40m telescope toward the Sgr B2 molecular cloud. Black cross indicates the location of G+0.693, where we have conducted our deep spectroscopic surveys with the IRAM 30m and Yebes 40m telescopes. Symbols show the location of the Sgr B2(N), (M) and DS massive star-forming regions.}
  \label{fig:map}
  \end{center}
  \end{figure}

As found for other GMCs (see Section \ref{sec:chemstudies} above), the COM molecular emission toward this cloud shows very low excitation temperatures (T$_{ex}$$\leq$15 K), despite the high kinetic temperatures of the gas \citep[$\sim$70-150 K;][]{zeng2018}. This is due to the low H$_2$ gas densities of a few 10$^4$ cm$^{-3}$ measured toward G+0.693 \citep{zeng2020}, which yields the sub-thermal excitation of the COM emission. As a result, only the lowest energy levels of molecules can be populated at such low T$_{ex}$ and hence, the number of rotational transitions present in the measured spectra is significantly smaller as compared to hotter sources such as hot cores and hot corinos. In addition, at such low T$_{ex}$ the peak emission of the observed COM spectra shifts towards lower frequencies, which are cleaner from the contribution from smaller and lighter molecules \citep{jimenez2014,jimenez2022a}. Therefore, despite the broad linewidths observed for the molecular emission in this cloud \citep[of $\sim$20 km s$^{-1}$][]{requena2006,zeng2018}, the levels of line blending and line confusion in the observed spectra of G+0.693 are low, making it an excellent laboratory to search for new COM interstellar species, and in particular those of prebiotic interest with complex rotational spectra.

\section{Prebiotic COMs in the Galactic Center}

\subsection{Discovery of precursors of Prebiotic Systems Chemistry in the ISM}

Early theories on prebiotic chemistry proposed that life emerged through metabolic-only or replicative-only biochemical systems. However, theories proposing a purely metabolic or genomic scenarios without compartmentalization encountered numerous theoretical and experimental problems \citep[see e.g.][for a review]{ruiz-mirazo2014}. In recent years, however, a new scenario called {\it Prebiotic Systems Chemistry} has emerged in the field and proposes that primitive Earth was a huge chemical reactor in which a high diversity of precursor components was available so that they could progressively and concurrently turned into primordial metabolic, self-replicating, and membrane-bounded sub-systems. 

\begin{figure}
  \begin{center}
  \includegraphics[scale=0.4,angle=0]{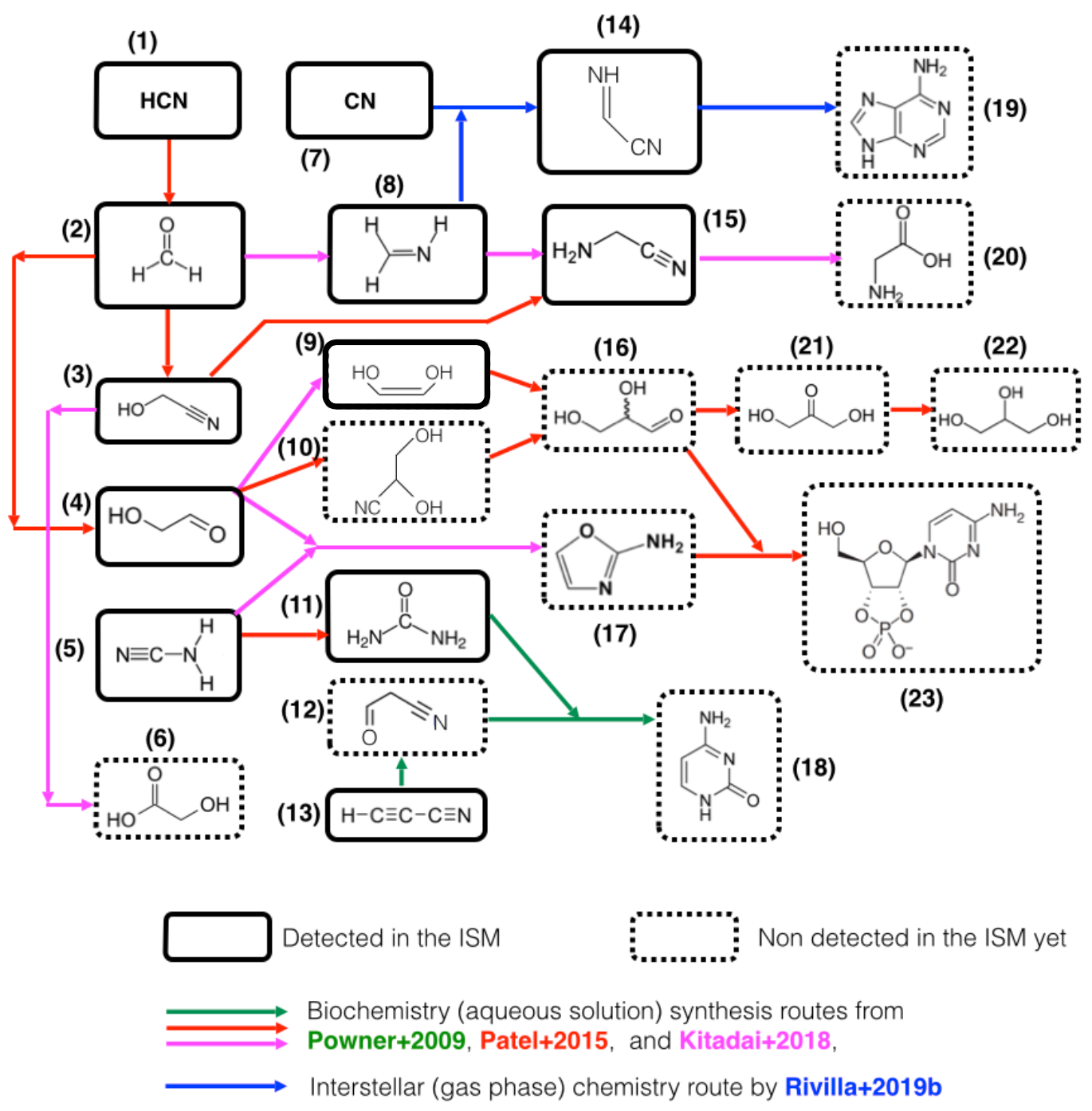}
  \caption{Example of precursors of ribonucleotides, amino acids and nucleobases proposed in the chemical schemes of \citet{powner2009}, \citet{patel2015} and \citet{kitadai2018}. This Figure has been updated from the work of \citet{jimenez2020} with the recent detection of 1,2-ethenediol \citep{rivilla2022a}. Solid boxes indicate molecules that have been detected in space, while black dotted boxes denote those species that remain undetected. Arrows show the possible synthesis pathways among the different molecular species within the scheme. The names of the molecules shown are: (1) Hydrogen Cyanide; (2) Formaldehyde; (3) Glycolonitrile; (4) Glycolaldehyde; (5) Cyanamide; (6) Glycolic acid; (7) Cyanide; (8) Methanimine; (9) 1,2-ethenediol; (10) Cyanohydrin; (11) Urea; (12) 3-Oxopropanenitrile; (13) Cyanoacetylene; (14) Cyanomethanimine; (15) Aminoacetonitrile; (16) Glyceraldehyde; (17) 2-amino-oxazole; (18) Cytosine; (19) Adenine; (20) Glycine; (21) Dihydroxyacetone (DHA); (22) Glycerol; (23) Beta-ribocytidine-2’,3’-cyclic phosphate (pyrimidine ribonucleotide).}
  \label{fig:prebiotic}
  \end{center}
  \end{figure}

Taking this scenario as a starting point, in 2018 we started a new project at the Center of Astrobiology (CAB) to try to understand whether key precursors in prebiotic systems chemistry could form in the ISM. To do this, we inspected the different chemical schemes proposed for the origin of the three properties of life: replication, metabolism and membrane compartments. Some examples are the works of \citet{powner2009}, \citet{patel2015},  \citet{becker2019}, and \citet{menor-salvan2009,menor-salvan2020}, which focus on the formation of ribonucleotides (the elemental units of RNA) under prebiotic conditions. In addition, the works of \citet{ruiz-mirazo2014} and \citet{kitadai2018} propose key species involved in proto-metabolic processes and precursors of proto-membranes. The common denominator in all these chemical schemes is that they typically start from simple molecules such as HCN, H$_2$CO, NH$_2$CN or glycolaldehyde, which are known to be abundant in the ISM. However, some others had not been identified in interstellar space up to 2018 (see Figure \ref{fig:prebiotic}) and hence, we started looking for them toward the chemically-rich G+0.693 molecular cloud \citep[][]{jimenez2020}.

\begin{figure}
  \includegraphics[scale=0.5,angle=0]{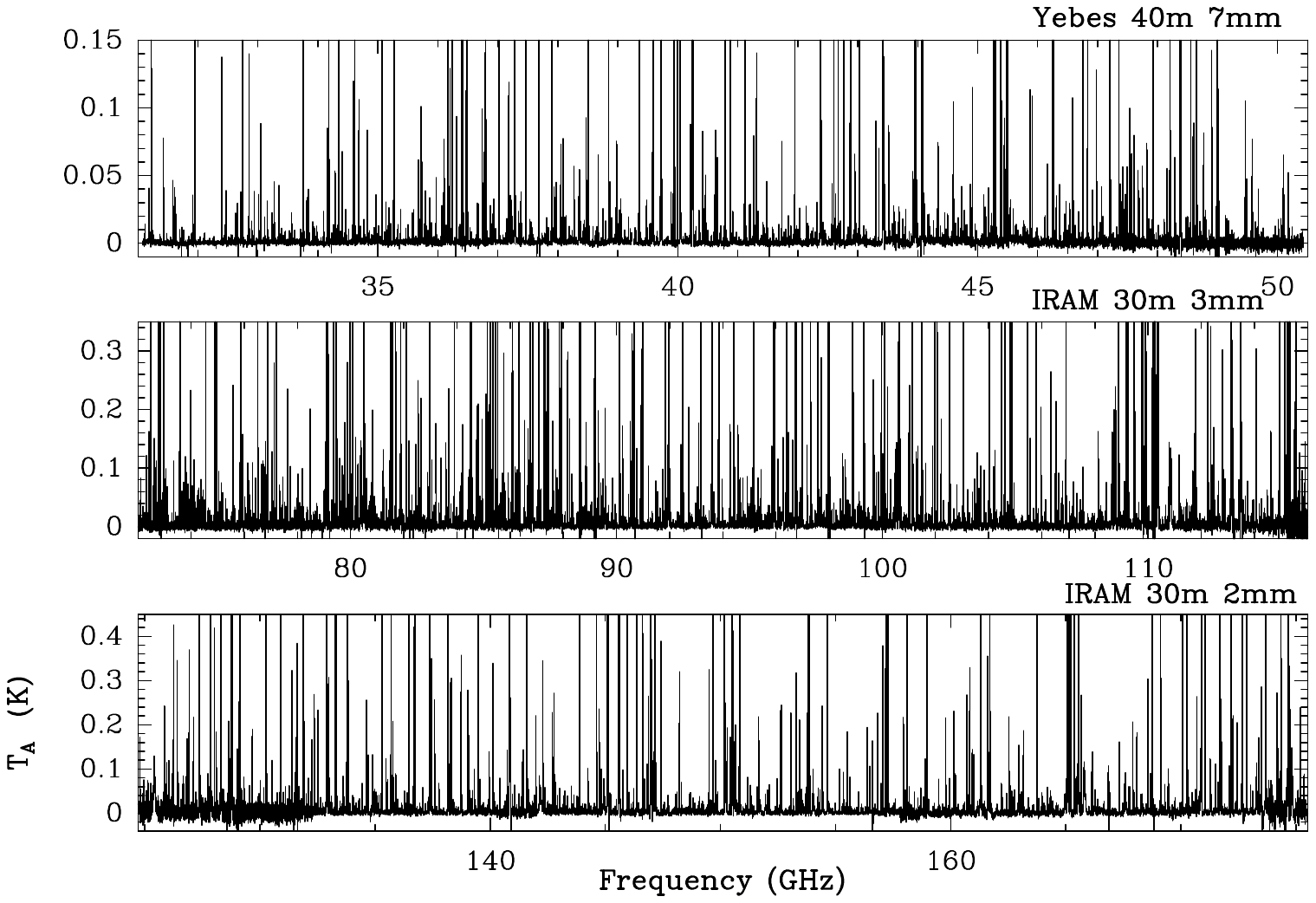}
  \caption{Frequency coverage of the spectroscopic surveys carried out at 7mm, 3mm, and 2mm toward the G+0.693 molecular cloud with the Yebes 40m and the IRAM 30m radiotelescopes. The observed spectra present a miriad of molecular rotational lines. For details on the surveys, see \citet{sanz-novo2023}.}
  \label{fig:survey}
  \end{figure}

In the past few years, we have collected unbiased, ultra-sensitive spectral surveys at 7mm, 3mm and 2mm toward G+0.693 using the Yebes 40m and IRAM 30m telescopes. The rms noise achieved in these surveys reach the sub-mK level at a velocity resolution of 1-1.5 km s$^{-1}$ and therefore, these observations represent the most sensitive surveys ever carried out toward a Galactic Center GMC. In Figure \ref{fig:survey}, we present the spectral coverage of the IRAM 30m and Yebes 40m spectral surveys at 7mm, 3mm and 2mm. The obtained spectra present a miriad of rotational molecular lines. Over 170 molecular species have been identified to date in our surveys, where more than 50\% of the detected species are COMs. 

Up to the submission of this proceeding chapter, we have reported the detection of 23\footnote{This number does not include urea (see Table \ref{tab:species}).} new molecular species in the ISM, 17 out of which are of prebiotic interest. In Table \ref{tab:species}, we report the prebiotic molecular species discovered to date toward the G+0.693 molecular cloud, together with their measured abundances. As shown in this Table, we have detected precursors of ribonucleotides, nucleobases, amino acids, carboxylic acids, sugars, proto-lipids and proto-proteins proposed in chemical schemes of prebiotic systems chemistry \citep{powner2009,patel2015,kitadai2018,becker2019,menor-salvan2020}. The measured abundances range from $\sim$5$\times$10$^{-12}$ to $\sim$5$\times$10$^{-10}$ with respect to H$_2$. We ignore much about the chemistry of these new species and new efforts have recently been carried out to understand their formation pathways and survivability under interstellar conditions (see Section \ref{sec:chemistry}). 

The majority of the detected prebiotic COMs likely form on the surface of dust grains \citep[see e.g.][]{rodriguez-almeida2021a,rodriguez-almeida2021b,rivilla2021a,jimenez2022b}. Only for a few cases, gas-phase reactions may be responsible for their formation \citep[see e.g.  C$_2$H$_3$NH$_2$ and PO$^+$;][]{zeng2021,rivilla2022b}. If these prebiotic compounds formed purely on grain surfaces, their fraction in interstellar ices would be $\sim$0.3-7 ppm with respect to water, assuming that all ices are sputtered into the gas phase\footnote{G+0.693 is believed to be affected by a $\sim$20 km s$^{-1}$-shock \citep[][]{rivilla2022b,sanz-novo2023}, which implies that the ice mantles have been fully sputtered from dust grains \citep[see][]{caselli1997,jimenez2008}.} and that the water ice abundance is $\sim$7.5$\times$10$^{-5}$ \citep[][]{whittet1991,mcclure2023}. This fraction is of the same order as those found in meteorites for some of these prebiotic compounds \citep{rivilla2021a}, which suggests that the ISM is an important source of prebiotic material that could have been delivered to Earth during the Late Heavy Bombardment period \citep{chyba1992}, contributing to the process of the origin of life \citep[e.g.][]{ruiz-mirazo2014,fiore2022}. 

\begin{table}[h!]
 \centering
 \caption{Summary of the prebiotic species discovered toward G+0.693 and their measured abundance}\label{tab:species}
 {\tablefont\begin{tabular}{@{\extracolsep{\fill}}ccccccc}
    \midrule
    Precursor of & Name & Formula & Abundance & Reference \\
    \midrule
    Ribonucleotides & Urea$^1$ & NH$_2$CONH$_2$ & 5.2$\times$10$^{-11}$ & \citet{jimenez2020,zeng2023} \\
    &   Hydroxylamine & NH$_2$OH &  2.1$\times$10$^{-10}$ & \citet{rivilla2020} \\
    Nucleobases & Z-Cyanomethanimine & Z-HNCHCN &  5.5$\times$10$^{-10}$ &  \citet{rivilla2019}; \citet{san-andres2024} \\
    & N-Cyanomethanimine & H$_2$CNCN &  2.1$\times$10$^{-11}$ &  \citet{san-andres2024} \\
    Amino acids & Vynil amine & C$_2$H$_3$NH$_2$ & 3.3$\times$10$^{-10}$ &  \citet{zeng2021} \\
    & Ethyl amine & C$_2$H$_5$NH$_2$ & 1.9$\times$10$^{-10}$ &  \citet{zeng2021} \\
    & Propargylimine & HCCCHNH & 1.8$\times$10$^{-10}$ & \citet{bizzocchi2020} \\
    & Allylimine & CH$_2$=CH-CH=NH & 4.0$\times$10$^{-11}$ & \citet[][]{alberton2023} \\
    & Cyanomidyl & HNCN & 9.1$\times$10$^{-11}$ & \citet{rivilla2021b} \\
    & Glycolamide & NH$_2$C(O)CH$_2$OH & 5.5$\times$10$^{-11}$ & \citet{rivilla2023} \\ 
    Carboxilic Acids & Carbonic Acid & HOCOOH &  4.7$\times$10$^{-11}$ & \citet{sanz-novo2023} \\
    Sugars & (Z)-1,2-ethenediol & (CHOH)$_2$ & 1.3$\times$10$^{-10}$ & \citet{rivilla2022a} \\
    Proto-proteins & Monothioformic acid & HC(O)SH & 1.0$\times$10$^{-10}$ & \citet{rodriguez-almeida2021a} \\
    & Ethyl isocyanate & C$_2$H$_5$NCO & (4.7-7.3)$\times$10$^{-11}$ & \citet{rodriguez-almeida2021b} \\
    Proto-lipids & Ethanolamine & NH$_2$CH$_2$CH$_2$OH & (0.9-1.4)$\times$10$^{-10}$ & 
    \citet{rivilla2021a} \\
    & Ga-n-propanol & Ga-n-C$_3$H$_7$OH & 4.1$\times$10$^{-10}$ & \citet{jimenez2022b} \\
    & Aa-n-propanol & Aa-n-C$_3$H$_7$OH & 2.5$\times$10$^{-10}$ & \citet{jimenez2022b} \\
    & Phosphorus monoxide Ion & PO$^+$ & 4.5$\times$10$^{-12}$ & \citet{rivilla2022b} \\
    \midrule
    \end{tabular}}
\tabnote{\textit{Note}: [1] This is the second detection of urea in the ISM. Interstellar urea was discovered by \citet{belloche2019}. }
\end{table}

\subsection{Formation and survivability of precursors of prebiotic systems chemistry}
\label{sec:chemistry}

The observations of these prebiotic species toward the G+0.693 molecular cloud has triggered the interest of several groups to understand their formation and survivability under interstellar conditions. By using quantum chemical calculations, formation mechanisms on the surface of dust grains have been investigated for urea \citep{slate2020} and HC(O)SH \citep{molpeceres2021}. For urea, the main formation routes on grain surfaces are found to be the radical-radical reaction NH$_2$CO + NH$_2$ $\rightarrow$ NH$_2$CONH$_2$ \citep[also proposed by][]{garrod2008}, and the addition of NH$_3$ to protonated isocyanic acid, HNCOH$^+$, followed by the tautomerization of the product that yields protonated urea \citep{slate2020}. For HC(O)SH, its trans form is efficiently produced on water ices by two subsequent hydrogenations of OCS. If trans-HC(O)SH is released into the gas phase, it can undergo isomerization reactions through quantum tunneling as described in Section \ref{sec:isomer} \citep[see also][]{garcia2022}. 

The survivability of these prebiotic species has also been studied both in the laboratory and using quantum chemical calculations. Recent laboratory experiments of the energetic processing of pure ices and water mixtures of 2-aminooxazole and urea have revealed that urea is  more resilient to destruction by the energetic processing with UV radiation and cosmic rays \citep{mate2021,herrero2022}. This naturally explains why urea has been detected toward the G+0.693 molecular cloud while 2-aminooxazole has not \citep{jimenez2020}. The destruction of the ribonucleotide precursor NH$_2$OH has also been investigated by \citet{molpeceres2023} using quantum chemical calculations. These authors have shown that the low abundance of NH$_2$OH measured toward G+0.693 is likely due to efficient hydrogen abstraction reactions on the surface of dust grains occurring at dust temperatures $\leq$20 K. Since dust in Galactic Center GMCs presents rather low temperatures \citep{rodriguez2000,etxaluze2013}, this is a viable mechanism for NH$_2$OH destruction in Galactic Center GMCs as well as in cold molecular clouds in the Galactic disk. This is consistent with the lack of detections of this molecule in star-forming regions across the Galaxy \citep{pulliam2012,mcguire2015,ligterink2018}.  

\subsection{Detection of high-energy isomers in G+0.693}
\label{sec:isomer}

One of the most surprising results obtained toward the G+0.693 molecular cloud, has been the discovery of high-energy isomers of several COMs under ISM conditions. This was unexpected because the relative energies between isomers for some of these species are of a few kcal mol$^{-1}$, which corresponds to energies $\geq$500 K. In the past, it was believed that isomerization transformations in the gas phase are unlikely to occur at the low temperatures of the ISM, because the energy barriers of such reactions are very high \citep{vazart2015}. For instance, the isomerization transformation between the [Z]-/[E]-cyanomethanimine isomers and between the cis-/trans-HCOOH and cis-/trans-HC(O)SH isomers present energy barriers $\geq$2500 K
\citep[][]{garcia2021,garcia2022}. However, as recently shown by these authors, isomerization processes can become viable under interstellar conditions thanks to ground-state quantum tunneling effects. Indeed, in this type of calculations a multidimensional small-curvature treatment of the quantum tunneling is essential because curvature effects can increase notably the importance of tunneling especially at low temperatures \citep[][]{garcia2021,garcia2022}. 
Several stereoisomers detected toward G+0.693
present isomeric abundance ratios that are well explained by quantum tunneling effects, allowing these systems to reach the thermodynamic equilibrium at moderately low temperatures. Some examples of this process can be found for imines such as cyanomethanimine, ethanimine and 2-propyn-1-imine \citep{garcia2021}, monothioformic acid \citep{garcia2022}, the Ga and Aa conformers of n-propanol \citep{jimenez2022b}, and the $cis-cis$ and $cis-trans$ isomers of carbonic acid \citep{sanz-novo2023}. Note that isomerization transformations have also been studied on the surface of dust grains via quantum chemical calculations and shown to be important for the observed isomeric ratios \citep[see e.g.][]{molpeceres2022}.

\subsection{Molecular abundance trends with increasing chemical complexity}

\begin{figure}
\vspace{-2cm}
  \begin{center}
  \includegraphics[scale=0.5,angle=0]{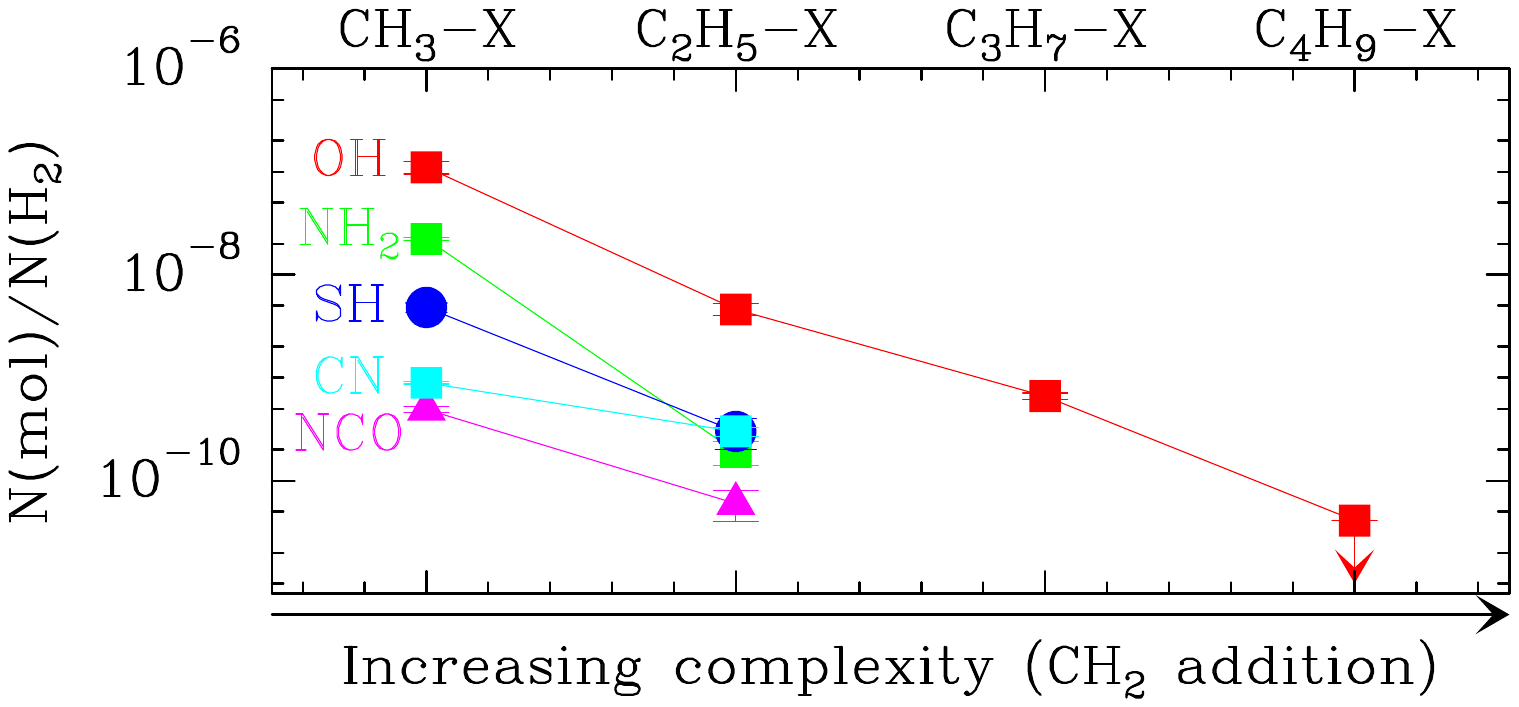}
\vspace{-2cm}
  \caption{Comparison of the abundances of molecular species of the form CH$_3$-X,  C$_2$H$_5$-X, C$_3$H$_7$-X, and C$_4$H$_9$-X measured toward G+0.693, with X the functional groups -OH (for alcohols; see red squares), -NH$_2$ (for amines; green squares), -SH (for thiols; blue circles), -CN (cyanides; light blue squares), and -NCO (for isocyanates; magenta triangles). Errors are shown for all species, but for some the errors are smaller than the size of the marker. Arrow indicates an upper limit. Chemical complexity increases from left to right. The abundances have been extracted from \citet{jimenez2022b} for alcohols, from \citet{zeng2021} for amines, from \citet{rodriguez-almeida2021a} for thiols, from \citet{zeng2018} for cyanides, and from \citet{rodriguez-almeida2021b} for isocyanates.}
  \label{fig:abundances}
  \end{center}
  \end{figure}

The unbiased spectroscopic surveys toward G+0.693 have provided the detection of a high number of COMs belonging to different chemical families. This allows us to measure the changes in abundance with increasing chemical complexity for the COMs belonging to the same  family. In Figure \ref{fig:abundances}, we report the abundances of the molecules of the form CH$_3$-X,  C$_2$H$_5$-X, C$_3$H$_7$-X, and C$_4$H$_9$-X measured toward G+0.693, with X the functional groups -OH (alcohols), -NH$_2$ (amines), -SH (thiols), -CN (cyanides), and -NCO (isocyanates). As reported in \citet{jimenez2022b}, the abundance of alcohols progressively decreases by roughly a factor of 10 with the addition of a CH$_2$ radical to their molecular structure. This is consistent with the chemical modelling results of \citet{charnley1995}, who assumed that the mantle abundances of higher homologues of methanol and ethanol decrease for increasing size by a factor of 10. As shown by these authors, these abundance ratios are preserved in the gas phase even at 10$^4$ yrs or 10$^5$ yrs after the injection of the ices from dust grains, which are consistent with the time-scales relevant for Galactic Center GMCs \citep{requena2006}. 

Besides alcohols, a similar trend is also found for thiols and isocyanates, for which the abundance of the  C$_2$H$_5$-X species decreases by one order of magnitude with respect to the abundance of the CH$_3$-X molecules (see Figure \ref{fig:abundances}). However, for the remaining chemical families, the trends are either steeper (as for amines with a CH$_3$-X/C$_2$H$_5$-X abundance ratio $\geq$100) or shallower (as for cyanides with a CH$_3$-X/C$_2$H$_5$-X abundance ratio $\sim$3). For amines, the steeper trend may be due to the fact that the chemistry of amines is more diverse than for alcohols, thiols and isocyanates, as demonstrated by the wide variety of amine compounds detected toward G+0.693 \citep{rivilla2020,zeng2021,rivilla2021a}. For cyanides, gas-phase chemistry may play an important role in the chemistry of these species, altering the initial grain mantle composition ratio injected into the gas phase \citep{zeng2018}.

%
%


\subsection{Other first detections toward G+0.693}

In addition to prebiotic molecules, the chemical richness of G+0.693 has enabled the discovery of other molecular species that, until recently, had remained elusive in the ISM. Examples are SiC$_2$ \citep{massalkhi2023}, isobutene \citep[(CH$_3$)$_2$C=CH2;][]{fatima2023}, HOCS$^+$ \citep{sanz-novo2024a}, HNSO \citep{sanz-novo2024b}, and the metal-sulfides NaS and MgS \citep{rey-montejo2024}. 
Interestingly, G+0.693 is a rich source in sulfur-bearing molecules, as demonstrated by the detection of new sulfur-bearing species such as HC(O)SH, HOCS$^+$, HNSO, NaS and MgS. This may be due to the high S/O abundance ratio found for this cloud as compared to other astronomical sources such as the Sgr B2(N) hot cores (see \citet{rodriguez-almeida2021a} and Jiménez-Serra, Codella \& Belloche 2024, Handbook of Astrochemistry, in press). This indicates that atomic sulfur may be little depleted in G+0.693 as a result of the presence of shocks in this cloud.

The detection of SiC$_2$ and of the metal-sulfide molecules NaS and MgS in the gas phase of G+0.693 requires the presence of low-velocity shocks \citep{massalkhi2023,rey-montejo2024}, because metals are known to be heavily depleted on dust grains \citep{Field1974,Konstantopoulou2024}. 
This is also the case of HNSO (the first interstellar molecule detected with N, S and O in its structure), which is likely formed on grain surfaces via the hydrogenation of the radical NSO and sputtered into the gas phase by shocks \citep{sanz-novo2024b}. By carrying out the modelling of the chemistry of HOCS$^+$, \citet{sanz-novo2024a} have found that its measured abundance is well explained by the processing of a 20 km s$^{-1}$-shock and an enhanced cosmic-ray ionization rate by more than a factor of 100 with respect to the standard value. In this model, HOCS$^+$ is formed in the gas phase by ion-neutral reactions after the release of OCS from dust grains \citep{sanz-novo2024a}. 


For isobutene, the dominant formation routes for this molecule remain vastly unexplored and hence detailed chemical modelling and new quantum chemical calculations are required to understand its chemistry in the ISM \citep{fatima2023}.  

\begin{table}[h!]
 \centering
 \caption{Summary of other new molecules detected toward G+0.693 and their measured abundance}\label{tab:nonprebiotic}
 {\tablefont\begin{tabular}{@{\extracolsep{\fill}}ccccccc}
    \midrule
    Chemical family & Name & Formula & Abundance & Reference \\
    \midrule
    S-bearing & Protonated Carbonyl Sulfide & HOCS$^+$ & 7.0$\times$10$^{-11}$ & \citet{sanz-novo2024a} \\
    & Thionylimide & HNSO & 6.0$\times$10$^{-10}$ & \citet{sanz-novo2024b} \\
     & Sodium sulfide & NaS & 3.7$\times$10$^{-13}$ & \citet{rey-montejo2024} \\
    & Magnesium sulfide & MgS & 4.5$\times$10$^{-13}$ & \citet{rey-montejo2024} \\
    Si-bearing & Silicon dicarbide & SiC$_2$ & 7.5$\times$10$^{-11}$ & \citet{massalkhi2023} \\
    C-bearing & Isobutene & (CH$_3$)$_2$C=CH2 & 1.9$\times$10$^{-9}$ & \citet[][]{fatima2023} \\
    \midrule
    \end{tabular}}
\end{table}

\section{Conclusions and perspectives}

The chemistry of GMCs in the Galactic Center is mainly governed by widespread low-velocity shocks and by an enhanced cosmic-ray ionization rate. This yields a rich chemistry in all sorts of molecular compounds from different chemical families. As a result, the GMCs located in the Galactic Center represent one of the richest molecular repositories in the Galaxy, including COMs and, in particular, compounds of prebiotic interest. 

Our unbiased, ultrasensitive and broadband spectral surveys carried out toward one of these Galactic Center GMCs (G+0.693) reveal that the ISM efficiently produces molecules considered to be precursors of ribonucleotides, nucleobases, sugars, proto-proteins, amino acis, carboxylic acids and proto-lipids. These species have also been found in meteorites and comets \citep[see e.g.][]{rivilla2021a,jimenez2022b,sanz-novo2023}, which suggests that essential compounds in prebiotic chemistry could have been formed in the ISM and subsequently incorporated into minor Solar-system bodies. One may think that the extreme physical conditions found in the ISM in the Galactic Center may make it a special place for prebiotic chemistry to emerge in space. However, the process of formation of a Solar-system is characterised by periods of energetic events \citep[e.g. energetic outbursts as those found in FU Ori stars or cosmic-ray rate enhancements produced by protostars and outflow shocks;][]{hartmann1996,padovani2021}, which affects the gas and dust in a scaled-down version of what is seen in the Galactic center. Detailed comparisons between the molecular abundances measured toward G+0.693 and toward other star-forming regions, both in the Galactic Center and in the Galactic disk, are undergoing (see Jiménez-Serra, Codella \& Belloche 2024, Handbook of Astrochemistry, in press, or López-Gallifa et al. in prep.).    

Finally, we note that the GMC G+0.693 studied by our group is only one of the many chemically-rich clouds present in the Galactic Center. The chemical complexity across the CMZ will be explored in the coming years at high-angular resolution thanks to the ALMA Large Program ACES (ALMA CMZ Exploration Survey; PI: Steve Longmore), providing dozens of new locations in the CMZ with equivalent levels of chemical complexity but whose measured spectra will have significantly narrower line profiles as a result of the filtering effect of the interferometer. However, the discovery of big prebiotic COMs such as sugars with three and four carbon atoms (as e.g. glyceraldehyde, dihydroxyacetone or erythrulose) may require absorption experiments against bright continuum background sources, which could be carried out with Band 1 of ALMA or with future instruments such as the Next Generation Very Large Array (ngVLA) and the Square Kilometre Array \citep[SKA;][]{jimenez2022a}.

\end{document}